# Epistemological Distinctions Between Science and History

Michael Courtney, PhD
Ballistics Testing Group, P.O. Box 24, West Point, NY 10996
Michael_Courtney@alum.mit.edu

Amy Courtney, PhD
Department of Physics, United States Military Academy, West Point, NY 10996
Amy_Courtney@post.harvard.edu

**Abstract:**
This article describes epistemological distinctions between science and history. Science investigates models of natural law using repeatable experiments as the ultimate arbiter. In contrast, history investigates past events by considering physical evidence, documentary evidence, and eyewitness testimony. Because questions of natural law are repeatably testable by any audience that exercises due experimental care, models of natural law are inherently more objective and testable with greater certainty than theories of past events.

Science was born from the human desire to acquire knowledge with greater certainty and objectivity than is available from "authoritative" sources and "expert" opinions. Since human opinion is fallible, a method was invented to provide greater objectivity through the testability of ideas.[1]

Science is the area of natural philosophy that seeks to describe natural laws and processes by demanding that hypothetical laws and descriptions of natural processes (scientific theories or models) are consistent with repeatable (or reproducible) experiments. Experimental repeatability allows scientific assertions to be tested with a level of objectivity and certainty unavailable in other areas of empirical knowledge.[2][3]

The two key components of this definition of science are the subject addressed (natural law) and the method of testing hypotheses (repeatable experiments).[4] Addressing different subjects or employing different methods may be valid and interesting intellectual endeavors, but they fall outside the scope of this definition of science.

The underlying idea behind science is that all physical processes are governed by a set of rules. Discovering these rules (or at least close approximations for specified circumstances) is the goal of scientific inquiry. Thus the questions science addresses are what physical processes are allowed to happen, under what circumstances does each process occur, and how can predictions be made regarding what will happen given sufficient information about the initial conditions.

The epistemological arbiter for scientific investigation of physical laws is repeatable (or reproducible) experiment.[5] Science demands that experiments that test (support or refute) hypothetical models be able to be reproduced a large number of times by any audience that exercises due experimental care. If the initial conditions are carefully and sufficiently controlled, the outcome of an experiment should be the same. (Quantum mechanics and stochastic processes introduce a probabilistic aspect to this, allowing for different possible outcomes, each with a well-defined probability of occurring in a large number of experimental trials. For example, smoking cigarettes might not always lead to cancer, but it certainly increases the probability of developing certain cancers.)

The demand of reproducible experiments gives science its objectivity, since any research group should be able to witness an experimental result if they take care to conduct their experiment under identical conditions to those reporting earlier results. This alone gives science a level of objectivity and certainty unavailable in other areas of empirical knowledge.[1]

---

[1] It is not our intent to make a distinction between reproducible observations and experiments. As long as the reproducible observations are testing predictions based on potentially falsifiable hypotheses regarding natural laws (or physical processes), observational science falls under the definition just as well as experimental science.

One of us (MC) was a member of a research team that computed thermal cyclotron absorption coefficients for binary stars so close together that mass from one star was falling into another star and creating lots of observable radiative effects. Our research was geared

Other definitions of science are possible, but they are less specific. Some might define science as what scientists do. In addition to begging the question, this definition has ambiguous boundaries, is considerably less objective, and is subject to changing over time.

Another common definition of science is the study of any idea or hypothesis by testing it against the available physical evidence. Forensic science is an example that falls under this definition. Forensic science is the application of science to analyze physical evidence to determine facts in matters of legal significance.

Most often, the facts in question are matters of "What happened?" rather than "What can happen and how?" As such, forensic science is more concerned with questions of history than questions of natural law, because it addresses the question of "What happened?" in a specific case. The distinction between forensic science and history is merely practical: forensic science anticipates a legal use of results.

From an epistemological viewpoint, history is distinct from science because it addresses different fundamental questions, and because it uses different methods. Historical questions are not testable by reproducible experiments beyond determining whether or not certain events are consistent with natural law. It is not possible to reproduce all the conditions with sufficient specificity without *a priori* knowledge of what happened. Historical hypotheses are commonly tested by considering three broad classes of evidence: eyewitness testimony, documentary evidence, and physical evidence.

Thus forensic science is really a historical discipline that focuses on analysis of the physical evidence to determine what happened as relevant to a legal proceeding. In the absence of corroborating documentary and eyewitness evidence, results of forensic science are not compelling in court. For example, the person collecting the evidence must testify to its source to establish its relevance, and the chain of custody must be documented throughout the process.

DNA matching provides another example of how documentary evidence must accompany forensic results to establish facts in court with the claimed level of certainty. The expert witness who conducted a DNA analysis often testifies to an extremely unlikely probability (1 in 100,000 to 1 in 10,000,000) that the DNA at the crime scene would match a random member of the population.

The impression in the mind of the jury is that it is extremely unlikely that the DNA came from anyone other than the defendant. However, unless it is established through documentary or eyewitness evidence that the defendant is not an identical twin, there is about a 1 in 300 chance that there exists an identical twin who could have been the source of DNA at the scene.

Thus, "forensic science" is a misnomer for the discipline that applies science to determine facts of legal significance, because it answers questions of historical fact using well-established physical laws rather than investigating hypothetical natural laws via repeatable experiments.

Why make an epistemological distinction between science and history? Certainly those applying science to historical questions should be trained in science to yield trustworthy analysis. However, there are important differences in potential testability [6] and in the relative levels of certainty.

Since both models and observations of natural law are testable with a very large number of reproducible experiments, they are subject to the potential of future testing, modification, and falsifiability [7] in ways that historical theories are not.[2] If the presuppositions that went into an

---

toward accurately predicting the radiative characteristics. There is a lot of research in various areas of astronomy either developing theory to predict radiative and absorptive characteristics of various stars and other astronomical objects or testing the theoretical predictions via careful observations. Some areas of cosmology are more concerned with the origin of the universe (which the above definition of science would classify as history). One of us (MC) took a graduate course at MIT from Alan Guth, who made important advances by adding the inflation model to the Big Bang theory. However, cosmology also tries to make testable predictions of future events, for example, whether or not the universe is open (will keep expanding indefinitely) or closed (will eventually contract under gravitation).

[2]Popper is cited here, but the point is consistent with many views of science including the intrumentalism favored by Bohr and the realism favored by Einstein.

experimental design or interpretations are found to be flawed long after the initial result, theories of natural law can be repeatably tested with revised presuppositions. For example, Newton's laws were found to be more limited in scope than originally thought centuries after they were first tested and verified.[8]  Models of natural law can be tested at a later time even if the original experimental data was falsified.  In contrast, if it is found that the initial crime scene investigators handled evidence dishonestly or carelessly, no amount of future testing is likely to secure a criminal conviction.

The death of eyewitnesses, the destruction of documents, and the degradation of physical evidence over time tends to slowly erode the ability to test historical theories. For example, the discovery of arsenic in Napoleon's hair hints at foul play, but we are less certain of the source of the hair sample than would be needed to gain a criminal conviction in court. Likewise, DNA testing might suggest a likelihood that Thomas Jefferson fathered children by a slave woman, but one would be hard pressed to win a paternity suit on the weight of evidence available in that specific case.

The ability to interpret historical evidence sometimes improves with technological advances, but the available evidence itself for testing historical theories only grows dimmer with time. In contrast, the ability to test natural laws steadily grows along with the scientific knowledge base and technological advances that can be leveraged for greater observational and interpretational ability.

The standard of testability for models of natural law is firm and immovable: reproducible experiment. In contrast, the standard of support for historical theories is much more flexible, depending on the amount of evidence that can reasonably be expected rather than on a rigid evidentiary standard universally applied to all scientific questions. The evidentiary standard for publishing an undisputed assertion in a history book about an event over 1000 years old is often a brief mention in a document more contemporary to the original event. The standard for historical events several hundred years old is usually a preponderance of documentary support. The standard for publishing news in a modern newspaper is usually confirmation by multiple sources. The standard for a criminal conviction is usually evidence sufficiently compelling to put the matter "beyond a reasonable doubt."

In conclusion, questions of natural physical laws can be answered more objectively than questions of history, because the results must be reproducible a very large number of times. Scientific theories can be revisited an indefinite number of times in the future as advances in the scientific knowledge base and available technology improve the observational abilities.

**About the Authors**
*Amy Courtney* currently serves on the Physics faculty of the United States Military Academy at West Point.  She earned a MS in Biomedical Engineering from Harvard University and a PhD in Medical Engineering and Medical Physics from a joint Harvard/MIT program.  She has published work in orthopedic biomechanics, biomedical engineering, ballistics, acoustics, medical physics, and traumatic brain injury.  She has taught Anatomy and Physiology as well as Physics.  She has served as a research scientist at the Cleveland Clinic and Western Carolina University, as well as on the Biomedical Engineering faculty of The Ohio State University.

*Michael Courtney* earned a PhD in experimental Physics from the Massachusetts Institute of Technology.  He has published work in theoretical astrophysics, theoretical and experimental atomic physics, chaos, ballistics, acoustics, medical physics, and traumatic brain injury.  He has also served as the Director of the Forensic Science Program at Western Carolina University and also been a Physics Professor, teaching Physics, Statistics, and Forensic Science.  Michael and his wife, Amy, founded the Ballistics Testing Group in 2001 to study incapacitation ballistics and the reconstruction of shooting events.